\documentclass[11pt,preprint]{aastex}
\shorttitle{SDSS J1001+5027 and SDSS J1206+4332}
\shortauthors{OGURI ET AL.}
\begin{document}
\title{Discovery of Two Gravitationally Lensed Quasars with Image
Separations \\
of 3 Arcseconds from the Sloan Digital Sky Survey}  
%
\author{
Masamune Oguri,\altaffilmark{1,2}
Naohisa Inada,\altaffilmark{3}
Joseph F. Hennawi,\altaffilmark{1,4,5}
Gordon T. Richards,\altaffilmark{1} \\
David E. Johnston,\altaffilmark{1}
Joshua A. Frieman,\altaffilmark{6,7}
Bartosz Pindor,\altaffilmark{8}
Michael A. Strauss,\altaffilmark{1} \\
Robert J. Brunner,\altaffilmark{9}
Robert H. Becker,\altaffilmark{10,11}
Francisco J. Castander,\altaffilmark{12}
Michael D. Gregg,\altaffilmark{10,11} \\
Patrick B. Hall,\altaffilmark{13}
Hans-Walter Rix,\altaffilmark{14}
Donald P. Schneider,\altaffilmark{15}
Neta A. Bahcall,\altaffilmark{1}\\
Jonathan Brinkmann,\altaffilmark{16}
and
Donald G. York\altaffilmark{6,17}
}
\altaffiltext{1}{Princeton University Observatory, Peyton Hall,
Princeton, NJ 08544.}
\altaffiltext{2}{Department of Physics, University of Tokyo, Hongo
7-3-1, Bunkyo-ku, Tokyo 113-0033, Japan.}
\altaffiltext{3}{Institute of Astronomy, University of Tokyo, 2-21-1
Osawa, Mitaka, Tokyo 181-8588, Japan.}
\altaffiltext{4}{Department of Astronomy, University of California at
Berkeley, 601 Campbell Hall, Berkeley, CA 94720-3411.} 
\altaffiltext{5}{Hubble Fellow}
\altaffiltext{6}{Astronomy and Astrophysics Department, University of
Chicago, 5640 South Ellis Avenue, Chicago, IL 60637.}
\altaffiltext{7}{Fermi National Accelerator Laboratory, P.O. Box
500, Batavia, IL 60510.}
\altaffiltext{8}{Canadian Institute for Theoretical Astrophysics,
University of Toronto, 60 St. George Street, Toronto, Ontario, M5S 3H8,
Canada.}
\altaffiltext{9}{Department of Astronomy, University of Illinois, 1002 W
Green Street, Urbana, IL 61801.}
\altaffiltext{10}{Department of Physics, University of California at
Davis, 1 Shields Avenue, Davis, CA 95616.}
\altaffiltext{11}{Institute of Geophysics and Planetary Physics,
Lawrence Livermore National Laboratory, L-413, 7000 East Aveneu,
Livermore, CA 94550.}
\altaffiltext{12}{Institut d'Estudis Espacials de Catalunya/CSIC,
Gran Capita 2-4, 08034 Barcelona, Spain.}
\altaffiltext{13}{Department of Physics \& Astronomy,
York University, 4700 Keele St., Toronto, ON M3J 1P3, Canada.}
\altaffiltext{14}{Max-Planck Institute for Astronomy, K\"onigstuhl
17, D-69117 Heidelberg, Germany.}
\altaffiltext{15}{Department of Astronomy and Astrophysics, Pennsylvania State
University, 525 Davey Laboratory, University Park, PA 16802.}
\altaffiltext{16}{Apache Point Observatory, P.O. Box 59, Sunspot, NM88349.}
\altaffiltext{17}{Enrico Fermi Institute, University of Chicago, 5640 South
Ellis Avenue, Chicago, IL 60637.}
%
%
\begin{abstract}
We report the discovery of two doubly-imaged quasars, SDSS
 J100128.61+502756.9 and SDSS J120629.65+433217.6, at redshifts of 
$1.838$ and $1.789$  and with image separations of $2\farcs86$ and
$2\farcs90$, respectively. The objects were selected as lens candidates
from the Sloan Digital Sky Survey (SDSS). Based on the identical nature
of the spectra of the two quasars in each pair and the identification of
the lens galaxies, we conclude that the objects are gravitational lenses.
The lenses are complicated; in both systems there are several galaxies
in the fields very close to the quasars, in addition to the lens
galaxies themselves. The lens modeling implies that these nearby
galaxies contribute significantly to the lens potentials. On larger
scales, we have detected an enhancement in the galaxy density near SDSS
J100128.61+502756.9. The number of lenses with image separation of $\sim
3''$ in the SDSS already exceeds the prediction of simple theoretical
models based on the standard Lambda-dominated cosmology and observed
velocity function of galaxies.  
\end{abstract}
 
 
\section{Introduction}

In strong gravitational lensing of quasars, the separations between
multiple images, $\theta$, is the most important observable linking
observations to theory. Since the image separation is determined by the
potential depth of the lens, the image 
separation distribution of lensed quasars offers a direct probe of the
hierarchical structure of the universe. For instance, normal galaxies
can produce strongly lensed quasars with image separations of 
$\sim 1''$, while  lenses with image separation $>10''$ can only be caused by
clusters of galaxies. About 70 of $\sim 1''$ lenses are known to
date\footnote{A summary of known lensed quasar systems is available 
on the CASTLES homepage (Kochanek, C.~S., Falco, E.~E., Impey, C.,
Lehar, J., McLeod, B., \& Rix, H.-W., 
http://cfa-www.harvard.edu/castles/)}, and there is one example of
a lensed quasar system in which the lens potential is dominated by that
of dark matter \citep{inada03b,oguri04a}. 

Among gravitationally lensed quasars, those with intermediate
image separations ($3''\lesssim \theta\lesssim 7''$ ) are of great
interest because they represent a link between small- and large-separation 
lenses. In the standard modeling procedure used to predict the 
distribution of image separations, assuming isothermal profiles  and an
a priori velocity function of galaxies,  lenses with image separations
$\gtrsim 3''$ are very rare, because even the largest early type
galaxies do not have Einstein radii this large. Thus the probability for
$\sim 7''$ lensing is almost negligible. However, while Q0957+561
\citep*[$\theta=6\farcs26$;][]{walsh79} is primarily lensed by a galaxy,
the image separation is boosted by the cluster in which the lensing
galaxy resides. This example implies that the environment of the lens
galaxy may significantly affect the distribution of image separations in
the $3''-7''$ range \citep*{keeton00,martel02}. In addition, a 
secondary mass along the line of sight could affect strong
lensing \citep*{wambsganss05}, and this also may enhance
the lensing probabilities in this image separation range. 
Finally, there is a predicted contribution in this range from
clusters; simple theoretical models that include
transition of the property of lenses at $\sim 10^{13}M_\odot$
\citep[e.g.,][]{oguri02} predict that large-separation lenses due to
clusters begin to dominate the total lensing probability. Therefore, the
overall lensing probability distribution for $\theta\gtrsim 3''$ is
predicted to depend on the interplay  of these two effects; the
environmental effects and the emergence of cluster lenses.  However, the
overall lensing probability at $\theta\gtrsim 3''$ is quite small, thus
a large number of quasars is needed to investigate the lensing
probability distribution. Indeed, even the current largest homogeneous
sample of lensed quasars \citep{myers03,browne03} contains only one lens
in this image separation range. 

In this paper, we present the discovery of two $\sim 3''$
gravitationally lensed quasars, SDSS J100128.61+502756.9 (hereafter SDSS
J1001+5027) and SDSS SDSS J120629.65+433217.6 (hereafter SDSS J1206+4332).
These gravitational lenses were identified from an ongoing lens search
using the data of the Sloan Digital Sky Survey
\citep[SDSS;][]{york00,stoughton02,abazajian03,abazajian04,abazajian05}.  
Currently the SDSS contains more than 50,000 spectroscopically
classified quasars; thus the SDSS provides the opportunity to construct
the largest homogeneous lens catalog in existence. Indeed, $13$ new
gravitationally lensed quasars have been found by using the SDSS
\citep[e.g.,][]{inada03a}. 
In this paper, we describe photometric and spectroscopic observations of
two new lens candidates and show that they are gravitational lenses.
We model the lens systems and discuss the environments of the lens
galaxies. We also compare the image separation distributions of lensed
quasars in the SDSS (though still very preliminary because of the
limited statistics) with a simple theoretical model.

This paper is organized as follows. In \S \ref{sec:sdss}, we briefly
describe our method of searching for lens candidates from the SDSS data.
Section \ref{sec:obs} presents the results of both photometric and
spectroscopic follow-up observations, and \S \ref{sec:model} shows
the result of lens modeling. Section \ref{sec:env} is devoted
to a discussion of the environments of the lens galaxies. We also
discuss the lensing probability distribution, which is shown in \S
\ref{sec:stat}. We summarize our results in \S \ref{sec:sum}.

\section{Selecting Candidates from the SDSS data}
\label{sec:sdss}

All gravitational lenses presented in this paper were selected as lens
candidates from the SDSS, which is a survey to image
$10^4{\rm deg^2}$ of the sky. The SDSS also conducts spectroscopy of
galaxies and quasars that are selected from the imaging data
\citep{eisenstein01,strauss02,richards02,blanton03}. A dedicated 
2.5-meter telescope at Apache Point Observatory (APO) is equipped with a
multi-CCD camera \citep{gunn98} with five optical broad bands centered at
$3551$, $4686$, $6166$, $7480$, and $8932${\,\AA} \citep{fukugita96,stoughton02}. 
The imaging data are automatically reduced by the photometric pipeline
\citep{lupton01}. The astrometric positions are accurate to about
$0\farcs1$ for sources brighter than $r=20.5$ \citep{pier03}, and the
photometric errors are typically less than 0.03 magnitude
\citep{hogg01,smith02,ivezic04}. The spectra cover $3800$--$9200${\,\AA}
at a resolution of $1800$--$2100$.  

We use spectroscopically classified quasars with $z>0.6$ to search for
gravitational lens candidates. SDSS J1001+5027 and SDSS J1206+4332 are
identified as lens candidates by our standard candidate selection
algorithm (N. Inada et al., in preparation). This algorithm is based on
the idea that the image of a quasar pair with a small separation appears
to be more extended than that of single quasar, and characterizes the
extent by the following SDSS image parameters: {\tt dev\_L} (the 
likelihood that the image of the object is fit by a de Vaucouleurs
profile), {\tt exp\_L} (the likelihood by an exponential disk), and {\tt
star\_L} (the likelihood by the point spread function).  
This algorithm has already found six new SDSS lenses
\citep{inada03a,inada05a,inada05b,pindor04,pindor05,oguri04b} as well as
all previously known gravitational lenses in the SDSS footprint. 
However, the possible second lensed components of the candidates we study in
this paper were also recognized as separate astronomical objects in the SDSS
imaging data because of the relatively large image separations
\citep{pindor03}.

\begin{figure}
\epsscale{0.7}
\plotone{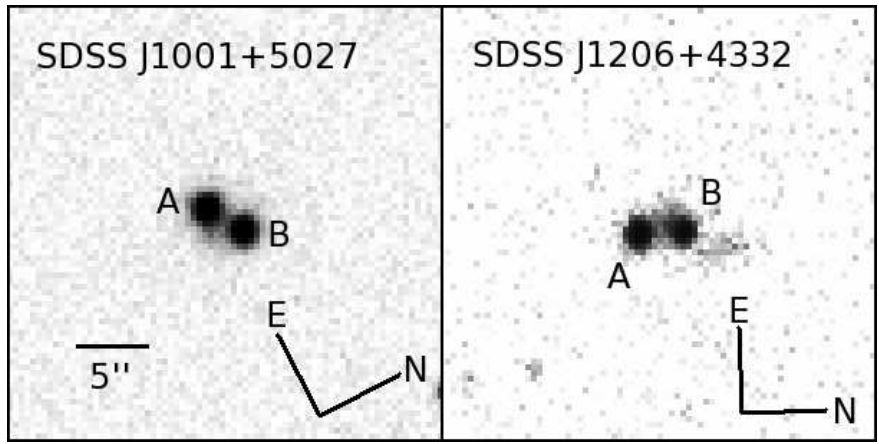}
\caption{The SDSS $i$-band images of SDSS~J1001+5027 ({\it left}) and
 SDSS~J1206+4332 ({\it right}). The image scale is 
 $0\farcs396$ ${\rm pixel^{-1}}$.  
\label{fig:sdss}}
\end{figure}

Figure \ref{fig:sdss} shows the SDSS $i$-band
images of the candidates. In both images, the seeing was $\sim
1\farcs2$, better than the typical seeing of SDSS images, $1\farcs35$
\citep{abazajian03}. The Figure clearly shows two stellar components in
each case; we designate the brighter ones as component A and the fainter
ones as component B. 
Table \ref{table:sdss} summarizes the result of the SDSS photometry
of the components. In each system the components have colors similar to
each other and the colors are consistent with that expected from
quasars; this makes them excellent lens candidates.

\section{Follow-up Observations}
\label{sec:obs}
Follow-up spectroscopic observations were conducted on 2003 November
20 and 2004 June 21 with the Double Imaging Spectrograph (DIS) on the
Astrophysical  Research Consortium (ARC) 3.5-m telescope at APO. We used
a $1\farcs5$ slit in low (blue) or medium (red) resolution mode. 
The resolution is about $2000$. The final spectra cover the wavelength range
of 3900{\,\AA} to  9400{\,\AA}. We aligned the slit along the direction
joining the two images; since the separations between the images are large
compared with the seeing size, extraction of the spectrum of each
component was straightforward. The data were reduced using standard
IRAF\footnote{IRAF is distributed by the National Optical Astronomy
Observatories, which are operated by the Association of Universities for
Research in Astronomy, Inc., under cooperative agreement with the
National Science Foundation.} reduction procedures. 

We also obtained deep images of the systems with the 8k mosaic CCD camera 
of the University of Hawaii 2.2-meter telescope (UH88) at Mauna Kea and
with the Seaver Prototype Imaging Camera (SPIcam) of the ARC 3.5-m telescope.
The UH88 imaging was conducted on 2004 May 23 and 25. We obtained
$VRI$-band images in the photometric nights; the exposure times were 360 
seconds for each candidate and each band. The image scale was $0\farcs232$ 
${\rm pixel^{-1}}$. We estimated the seeing as $\sim 0\farcs8$, better than
the SDSS seeing size of 
$\sim 1\farcs2$. The ARC $z$-band imaging was conducted on 2004
January 13. The exposure was 300 seconds, the seeing was $\sim 1\farcs0$, and
the image scale was $0\farcs282$ ${\rm pixel^{-1}}$. Each frame was
bias-subtracted and flat-field corrected. The magnitudes in the UH88 images
were calibrated by the standard star PG 1528+062
\citep{landolt92}. Astrometry and photometry of the UH88 images are
summarized in Table \ref{table:uh88}.   

\subsection{SDSS J1001+5027}

\begin{figure}
\epsscale{0.6}
\plotone{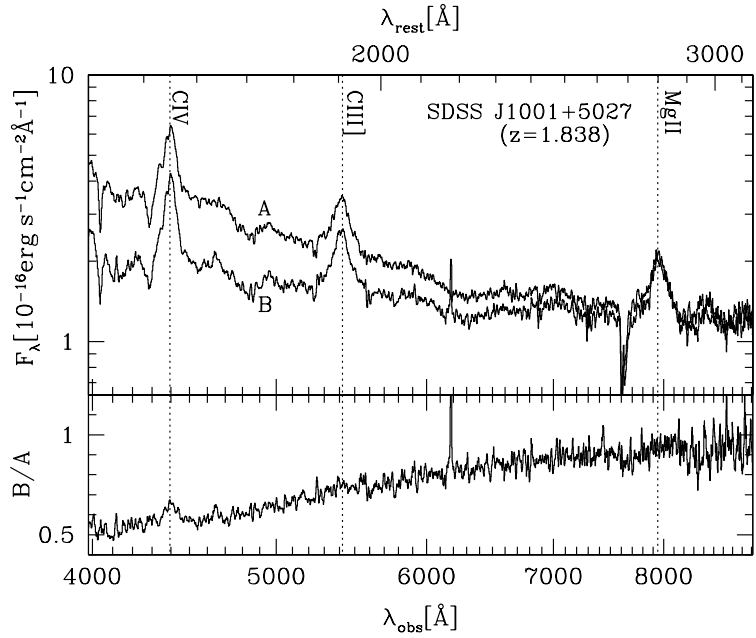}
\caption{ARC 3.5-m spectra of SDSS~J1001+5027 components A and B. The exposure
 time was 900 seconds. Both components have the same redshift, $z=1.838$.
 The strong absorption at $\sim 7600${\,\AA} is 
 atmospheric. The feature at $\sim 6200${\,\AA} is a bad column.
 The spectrum is smoothed by a 3 pixel boxcar.
 The ratio of the spectra as a function of wavelength 
 is shown in the bottom panel. B is evidently reddened.
\label{fig:spec1001}}
\end{figure}

\begin{figure}
\epsscale{0.8}
\plotone{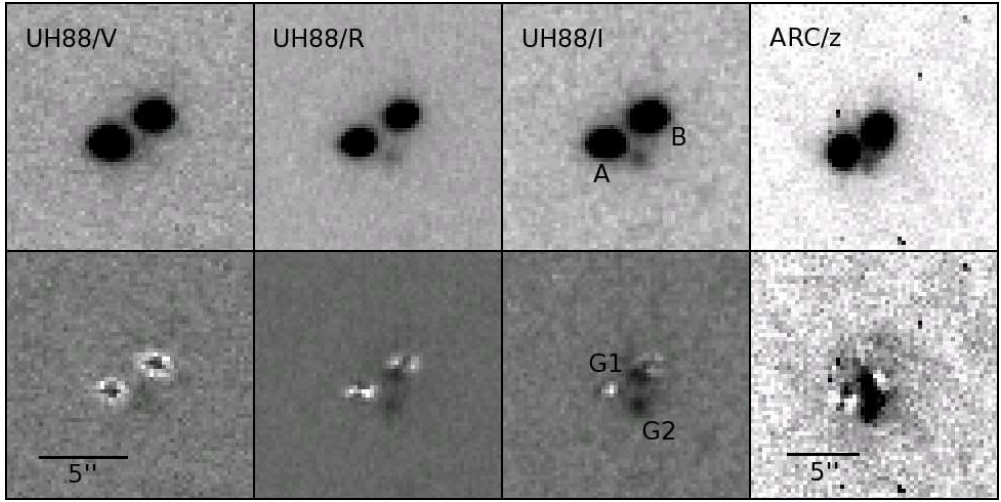} 
\caption{UH88 and ARC 3.5-m images of SDSS~J1001+5027. Lower panels are same
 as upper panels, but after subtracting two point source quasar
 components. The image scales are $0\farcs232$ ${\rm pixel^{-1}}$ (UH88) and 
 $0\farcs282$ ${\rm pixel^{-1}}$ (ARC), and the seeings were $\sim 0\farcs8$ 
 (UH88) and $\sim 1\farcs0$ (ARC).  North is up and East is left in all
 panels. We find two galaxies in the residual images shown in the lower
 panels, which are labeled G1 and G2.
 \label{fig:uh_1001}}
\end{figure}

Spectra and images of this object are shown in Figures
\ref{fig:spec1001} and \ref{fig:uh_1001}, respectively. In Figure
\ref{fig:spec1001}, both components show \ion{C}{4}, \ion{C}{3]}, and
\ion{Mg}{2} emission lines redshifted by $z=1.838$. It is worth
noting that the flux ratios (B/A) of the emission lines, particularly
the \ion{C}{4} emission lines, are larger than that of the continuum;
the equivalent widths of \ion{C}{4} emission lines are 56{\,\AA} and
76{\,\AA} for A and B, respectively. Such difference can be caused 
by the difference of the emission regions of continuum and broad
emission lines combined with microlensing by stars. 
We find two extended
objects near the quasar images (see Figure \ref{fig:uh_1001}; the image
separation is $2\farcs86$). When subtracting a quasar component, we used
a nearby star as a point-spread function (PSF) template.  The objects,
denoted by G1 and G2, have colors consistent with
those of early-type galaxies at $0.2\lesssim z\lesssim 0.5$ 
\citep*{fukugita95}. Since galaxy G1 is nearly
colinear with the two quasar components, it is likely that G1 is the
main contributor to the lens potential. However, galaxy G2 is
also quite close to the lens system, and could affect the lens
potential significantly. We note that component B could be
reddened by the lens galaxy G1. 

\subsection{SDSS J1206+4332}

\begin{figure}
\epsscale{0.6}
\plotone{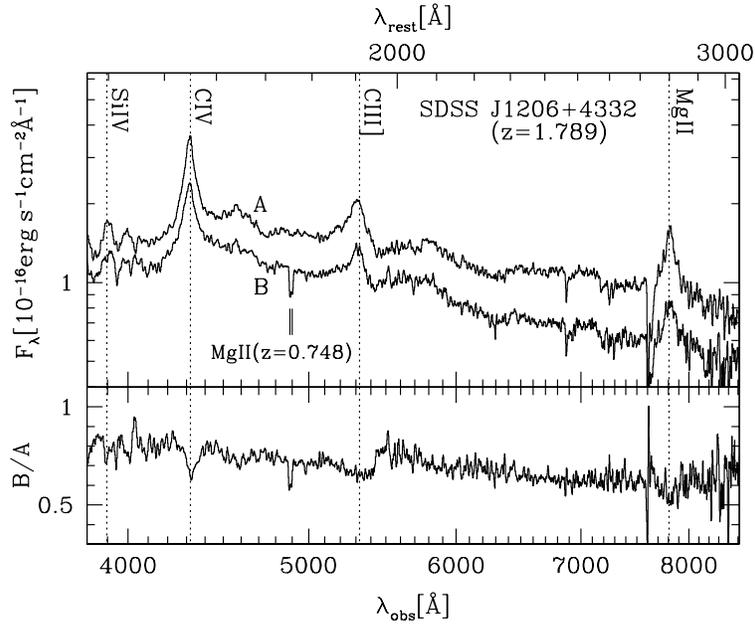}
\caption{ARC 3.5-m spectra of components A and B of SDSS~J1206+4332. The
 exposure time was 1800 seconds. The redshifts of both components are the same,
 $z=1.789$. The strong absorption at $\sim 7600${\,\AA} is
 atmospheric. The spectrum is smoothed by a 3 pixel boxcar. 
 There is strong \ion{Mg}{2} absorption at $\sim 4900${\,\AA} (corresponds
 to an absorber at $z=0.748$) in the spectrum of component B. The ratio
 of the spectra as a function of wavelength is shown in the bottom panel.
 \label{fig:spec1206}} 
\end{figure}

\begin{figure}
\epsscale{0.6}
\plotone{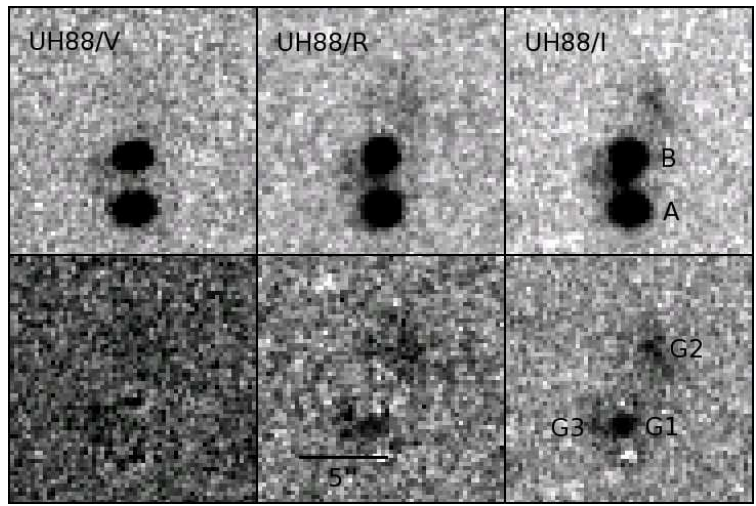}
\caption{UH88 images of SDSS~J1206+4332. Lower panels show images
 after subtracting the two quasar components. North is up and East is
 left in all panels. We find three galaxies in the  residual images
 shown in the lower panels, which are labeled G1, G2, and G3.
 \label{fig:uh_1206}}
\end{figure}

The spectra of the two components shown in Figure \ref{fig:spec1206} show 
\ion{Si}{4}, \ion{C}{4}, \ion{C}{3]}, and \ion{Mg}{2} emission
lines at the same wavelengths, which supports the idea that this is a 
gravitationally lensed quasar at $z=1.789$. However, the flux ratios (B/A)
of the emission lines are slightly smaller than that of the continuum;
the equivalent widths of \ion{C}{4} emission lines are 69{\,\AA} and
53{\,\AA} for A and B, respectively. The images in Figure
\ref{fig:uh_1206} clearly reveal the lensing galaxy G1 as well as two
quasar components separated by $2\farcs90$, further supporting the
lensing hypothesis. We also found other galaxies G2 and G3 near
component B. These galaxies, particularly G2, may contribute to the lens
potential to some extent. Indeed, both G1 and G2 are quite red (see Table
\ref{table:uh88}) and are
consistent with being high-redshift ($z\gtrsim 0.7$) early-type galaxies
\citep{fukugita95}, while G3 is blue and thus may be a chance
superposition of a local galaxy. There is strong \ion{Mg}{2}
absorption (equivalent width $>2${\,\AA}) at $\sim 4900${\,\AA} in the
spectrum of component B.  The redshift of the absorber is $z=0.748$,
consistent with the color of G1. Therefore the absorber may be
associated with the lensing galaxy G1.  
 
\section{Lens Modeling}
\label{sec:model}
Although the lens systems appear quite complex, we will first attempt to
model the lens systems with simple models. Specifically, we try
models that 
describe G1 by a Singular Isothermal Ellipsoid (SIE), or Singular
Isothermal Sphere (SIS) plus external shear. Even these simple 
models, however, have eight parameters (the galaxy position $x_{\rm g}$
and $y_{\rm g}$, the Einstein ring radius $R_{\rm E}$, the ellipticity
$e$ or shear $\gamma$, the position angle  $\theta_e$ or
$\theta_\gamma$, the source position $x_{\rm s}$ and $y_{\rm s}$, and
the flux of the quasar $f$) that is equal to the number of observational
constraints from the UH88 imaging data (the image positions, the galaxy
position, and fluxes of the images; see Table \ref{table:uh88}). 
Thus there are no degrees of freedom and in usual cases we will be able
to find models that perfectly reproduce the observables. To fit the
models, we use standard lens modeling techniques as 
implemented in the {\it lensmodel} software \citep{keeton01}. 

SDSS J1001+5027 is well fitted by both the SIS plus shear and SIE models.
The resulting fitting parameters are shown in Table \ref{table:model}.
The ellipticity $e=0.25$ in the SIE model is similar to that of the light,
though the position angle $\theta_e=10.9$ (measured East of North) is
quite different from that of the light ($\sim -60^\circ$) measured from
the UH88 image. In general the position angles of the 
light and lens models are aligned \citep*{keeton98}; therefore this
result suggests that the external field, rather than the galaxy G1, is
responsible for the quadrupole moment of the lens potential. 
The position angles in the models are rather close to the direction to G2.
Since in general the position angle of external shear gives an idea of the
direction to a main perturber, it seems that G2 significantly
affects the lens models. We also predict the time delay as 
$\Delta t\sim 45 h^{-1}{\rm day}$ (A leads B), assuming a lens
redshift of $z=0.3$.   

Fitting SDSS J1206+4332 by either SIS plus shear or SIE failed; the models
yielded large chi-squares $\chi^2>2$ with no degrees of freedom, and the
models required unnaturally  large $e$ or $\gamma$. We 
also tried a SIE plus shear model, but the resulting fit was similarly
poor. This implies that the lens system is too complicated to be
described by such simple models. Thus we add G2, which is modeled
by a SIS, as well as galaxy G1 modeled by SIE, in order to make the
model more realistic. We derive best-fit models by changing the value 
of the Einstein radius of G2, $R_{\rm E}$(G2). We find that this
``SIE$+$G2'' model fits 
the data well when $R_{\rm E}{\rm (G2)}\lesssim 1''$,
although the maximum value is slightly smaller than that inferred from
the Faber-Jackson relation ($R_{\rm E}{\rm (G2)}/R_{\rm E}{\rm
(G1)}=0.92$ from the $R$-band flux ratio, assuming G1 and G2 are at the
same redshift). As a specific example, we show the best fit
parameters for $R_{\rm E}{\rm (G2)}=1''$ in Table
\ref{table:model}. The position angle of galaxy G1 in
the model ($\theta_e=-89.3$) is in good agreement with that
observed. The time delay is derived to be $\Delta t = 92.6 h^{-1}{\rm
day}$  (A leads B), assuming the strong \ion{Mg}{2} absorption system
at $z=0.748$ is associated with the lens galaxy. Even if we decrease the
value of $R_{\rm E}$(G2) up to $0\farcs1$, the time delay is affected
only moderately; the time delay is predicted to be $104.4 h^{-1}{\rm
day}$ when $R_{\rm E}{\rm (G2)}=0\farcs1$.   

In summary, lens modeling has revealed that neither system is
simple. In particular, the secondary galaxies may play an important role
in both lenses. 

\section{Lens Galaxy Environments}
\label{sec:env}

Lens galaxies of lensed quasars, particularly in systems with relatively
large image separations ($\gtrsim 3''$), commonly lie in groups or clusters
\citep{keeton00,faure04}. Such ``compound'' lens systems include Q
0957+561 \citep{walsh79}, PG 1115+080 \citep{weymann80}, MG 2016+112
\citep{laurence84}, RX J0911+0551 \citep{bade97}, MG 0751+2716
\citep{tonry99}, SDSS J0903+5028 \citep{johnston03}, CLASS B1608+656
\citep{fassnacht04}, B2108+213 \citep{mckean05}, and HE 0435-1223
\citep{morgan05}. If dense environments of the lens galaxies are 
common, then they could affect strong lens studies in several ways
\citep{keeton04}. On the theoretical side, estimation of environmental
effects remains controversial:  While \citet{keeton00} and \citet{holder03}
argued that the large fraction of lens systems should lie in dense
environments and thus the environmental effects are significant,
\citet{dalal05} estimated using a halo occupation distribution that the
typical values for the external convergence and shear are quite small. 
Although we have already seen in \S \ref{sec:obs} and \S \ref{sec:model} 
that SDSS J1001+5027 and SDSS J1206+4332 
have additional galaxies that likely affect the lens potential, it is
important to check the larger field for hints of groups or clusters. 

\begin{figure}
\epsscale{0.9}
\plotone{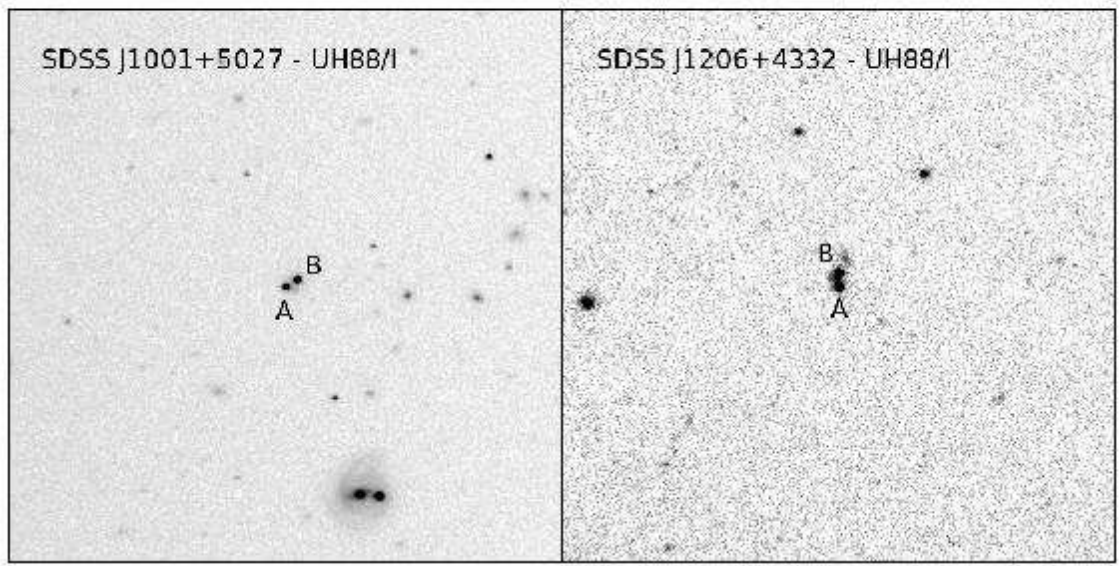}
\caption{The $2'\times 2'$ $I$-band fields around the lens system
 SDSS~J1001+5027 ({\it left}) and SDSS~J1206+4332 ({\it right}) taken at
 UH88. The exposure times were both 360 seconds. North is up and East is
 left in both panels. 
 \label{fig:fields}}
\end{figure}

Figure \ref{fig:fields} shows the fields around the lens systems
obtained at UH88. We find many galaxies around SDSS J1001+5027,
indicating that there may be a group or cluster along the line of sight. 
For SDSS J1206+4332, we cannot see any noticeable enhancement of the
number of galaxies around the lens system, though there are several
faint galaxies near the lens system.  

\begin{figure}
\epsscale{0.7}
\plotone{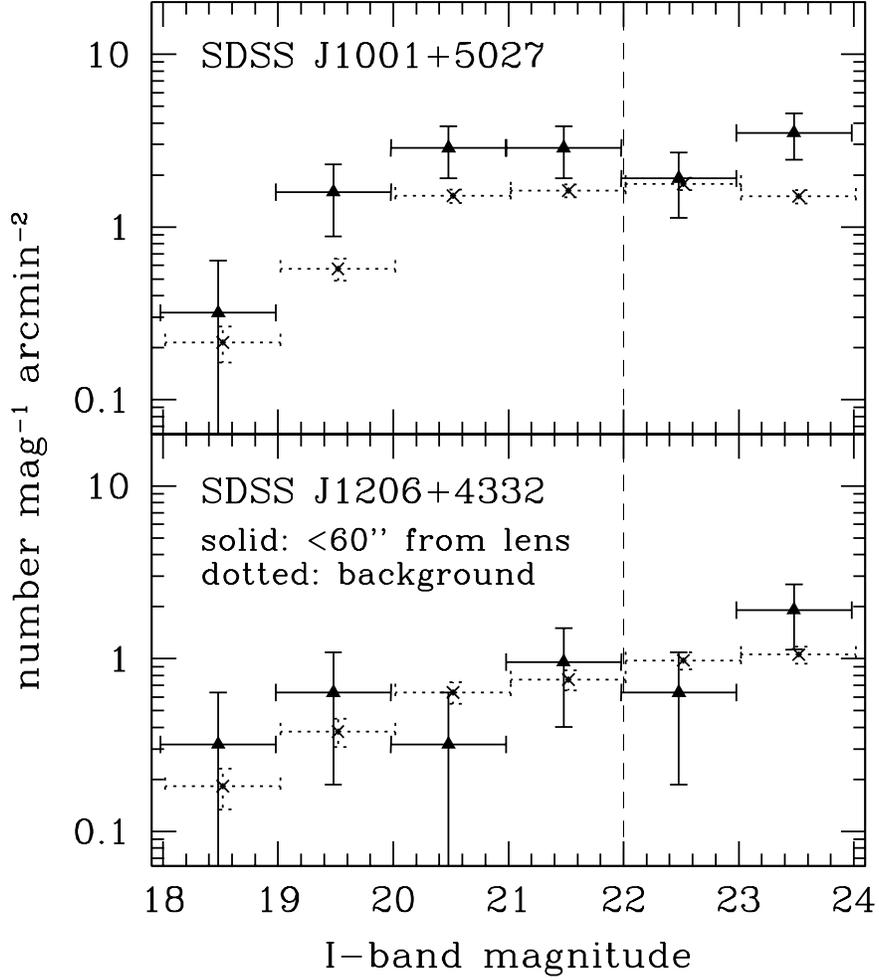}
\caption{$I$-band galaxy number counts of fields within $60''$ of
 the lens systems ({\it filled triangles with solid errorbars}) as well
 as background counts ({\it crosses with dotted errorbars}).  
 The errorbars include Poisson errors only. The vertical dashed lines
 indicate the approximate limit of the star/galaxy
 separation.\label{fig:count}} 
\end{figure}

To explore the environments further, we derived the number densities of
galaxies around each lens system using the UH88 images. We perform object
identifications 
using the Source Extractor algorithm \citep[SExtractor;][]{bertin96}. 
We define galaxies as objects with SExtractor parameter {\tt
CLASS\_STAR} smaller than 0.6 in the $I$ band image. Note that this
star/galaxy separation criterion is successful only for objects with 
$I \lesssim 22$. We derive the number densities of galaxies as a function
of $I$-band magnitudes of the galaxies, which are shown in Figure
\ref{fig:count}. 
In the Figure, we compare the galaxy number densities within $60''$ of 
each lens system with those of the background, estimated from galaxies 
more than $60''$ away in each image after excluding the region near
bright stars. The total areas of  the backgrounds are thus 87.0 
${\rm arcmin^2}$ and 79.8 ${\rm arcmin^2}$ for SDSS J1001+5027 and SDSS
J1206+4332, respectively.  We find an enhancement of galaxy number
densities for SDSS J1001+5027; 
the number densities of three magnitude bins show excesses by more than
$1\sigma$. On the other hand, the galaxy number count around SDSS
J1226+4332 is consistent with that of the background.

\begin{figure}
\epsscale{0.9}
\plotone{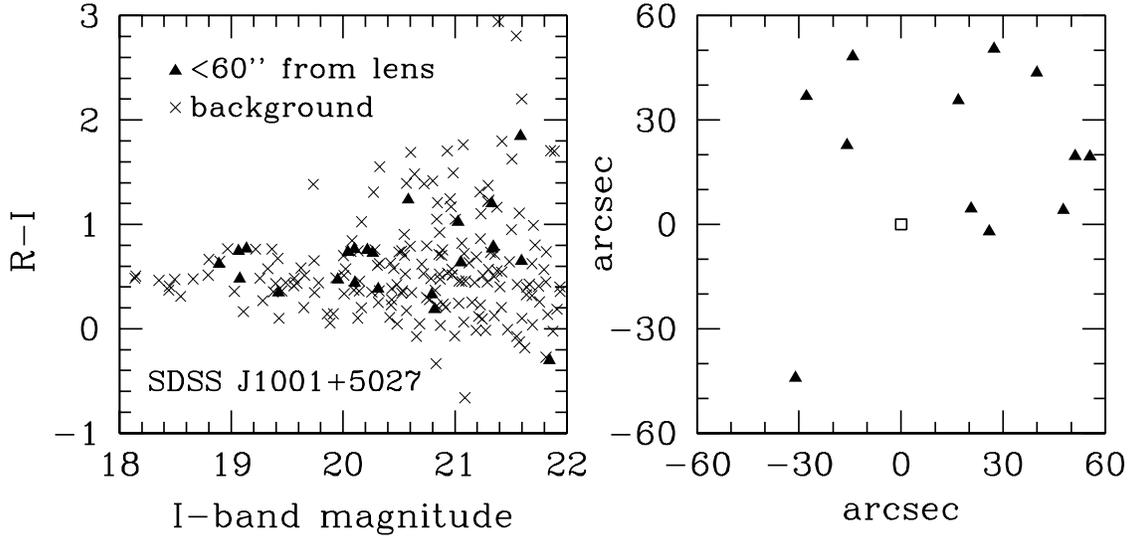}
\caption{{\it Left:} Color-magnitude diagrams of galaxies in the field
 within $60''$ of SDSS J1001+5027 ({\it filled triangles}) and the
 background ({\it crosses}). {\it Right:} The distribution of galaxies
 at $z\gtrsim0.2$ ({\it filled triangles}) around the lens system SDSS
 J1001+5027 ({\it open square}). North is up and East is left.
 \label{fig:colmag}}
\end{figure}

Figure \ref{fig:colmag} shows a color-magnitude diagram of galaxies near
SDSS J1001+5027 to study the origin of the enhancement of galaxy number
densities. We find a weak ridge line at $R-I\sim 0.75$, suggesting the
existence of a group or a cluster at $z\sim 0.2$, which is consistent
with that of lens galaxies G1 and G2 estimated from their colors. 
We also plot the distribution of galaxies at $z\gtrsim0.2$, by making
color cuts; $V-R>0.9$, $R-I>0.7$, and $I<22$. We find that the galaxies are
clustered North and West of the lens system, rather than distributed
homogeneously (see also Figure \ref{fig:fields}). Thus the lens galaxy
may lie in a group or a cluster that is located to the north-west of the
lens; spectroscopic identifications of these galaxies as well as the
lens galaxy should be undertaken.  

\section{Is the number of 3'' lenses consistent with theory?}
\label{sec:stat}

Thus far, 13 gravitationally lensed quasars have been discovered using the
SDSS. In addition, we have recovered several previously known lensed
quasar systems. Table \ref{table:lens} summarizes the current status of
our lens search in the SDSS. Note that limitations of follow-up time
have forced us to focus on the SDSS spectroscopic sample of
quasars\footnote{The photometric quasar sample \citep{richards04}  
has an order of magnitude more quasars than the spectroscopic sample,
and thus is expected to contain many more lenses.}. Thus the list does
not contain lenses that do not have SDSS spectra \citep[e.g., APM
08279+5255; see][]{pindor03}. The Table does not contain 
SDSS J1402+6321 \citep{bolton05} either, since the redshift ($z=0.48$) 
is below our criterion for the quasar sample (see discussion
below). Among gravitational lenses in the Table, SDSS J1004+4112 and Q
0957+561 were selected by 
searching around each quasar for stellar objects that have similar
colors as the quasar itself \citep{oguri04a,hennawi04}, and the rest
were successfully selected by our standard candidate selection algorithm
(see \S \ref{sec:sdss}) though the first identifications  of some of the
candidates were made by different algorithms. 
An exception is SDSS J0903+5028 \citep{johnston03} which was
targeted as a luminous red galaxy \citep{eisenstein01} because of the
bright lens galaxy; however, we include this system in the statistical
analysis below, since the quasar would have been targeted as a
high-redshift quasar if it had not been obscured by the foreground lens
galaxy \citep{richards02}. The Table shows that we have discovered a
relatively large number of $\sim 3''$ lenses. We note that our current
lens sample shown in Table \ref{table:lens} is quite
incomplete\footnote{We note that our follow-up observations are biased
against discovering lensed quasars with separations $<2''$ because of
poor seeing in many follow-up observations. This mostly explains the
large discrepancy between theory and observation at the image separation
range.}, and that future follow-up observations would increase the
number of lenses even in the current quasar sample.  Therefore we now
calculate the expected lensing probability in the SDSS quasar sample and
compare it with the number of  $\sim 3''$ lenses in the sample. In particular,
we neglect the contribution of lens galaxy environments, as has
normally been done, to see whether the assumption is still valid or not. 
In computing the lensing probability, we assume a spatially flat
universe ($\Omega_M+\Omega_\Lambda=1$). 

We compute the lensing probability distribution along the lines described by
\citet*{turner84}. The lens galaxies are modeled by SIS 
$\rho(r)=\sigma^2/(2\pi G r^2)$, where $\sigma$ is the
velocity dispersion of the lens galaxy. We adopt the velocity function of
early-type galaxies determined from $\sim 30,000$ SDSS galaxies 
at $0.01<z<0.3$ \citep{sheth03,bernardi05,mitchell05}, and neglect
the redshift evolution (i.e., we assume that
the velocity function is constant in comoving units). 
The use of the velocity function of early-type galaxies is sufficient for
our calculation because at $\theta\gtrsim 3''$ lensing by early-type
galaxies is dominant \citep{turner84}. We need also to incorporate
the selection function of the SDSS lens search; which we do in a
preliminary way by making the following assumptions.
First, we use the magnification factor of the brighter image, 
$\mu_{\rm bright}=(\theta_{\rm E}/\theta_{\rm S})+1$, where 
$\theta_{\rm E}$ is the Einstein radius and $\theta_{\rm S}$ is the
position of the source relative to the lens galaxy, to compute the
magnification bias, because at $\theta\gtrsim 3''$ two lensed components
are well separated in the SDSS data. Although we may have an alternative
choice of using the magnification factor of the fainter image, we adopt
$\mu_{\rm bright}$ to calculate the upper limit of the lensing
probability; thus the actual lensing rate at $\theta\gtrsim 3''$ might
be smaller than our calculation\footnote{Note that at smaller separation
($\theta\sim 1''$) our computation is expected to underestimate the
lensing probability because at such small separations lensed components
are not decomposed and thus the magnification factor of the two images
$\mu_{\rm total}=2\theta_{\rm E}/\theta_{\rm S}$ is more appropriate.}.
Next, for the limiting bright-to-faint flux ratio we assume $f_{\rm
max}=10$; this is justified because all lenses in Table \ref{table:lens}
satisfy this condition.  Lensing probabilities calculated
with this selection function represent an upper limit at relatively
large image separation $\theta\sim 3''$.

To calculate the expected number of lensed quasars in the current SDSS
quasar sample, we need the luminosity function of quasars as well as 
the redshift and magnitude distributions of the quasar sample. We adopt
a sample of $\sim 47,000$ quasars at $0.6<z<4.0$. The sample is
constructed in the same way as in \citet{oguri04a}. We did not search
for lensed quasars at $z<0.6$ as low-redshift quasars are intrinsically
extended, and we do not use high-redshift quasar ($z>4.0$) sample because 
it contains a significant fraction of objects whose spectra were
misidentified by the SDSS spectroscopic pipeline \citep[c.f., discussion
in][]{schneider03}. The redshift 
distribution is similar to that shown in \citet{oguri04a}. For the
luminosity function of quasars, we use what is called LF1 in
\citet{oguri04a}, which has a faint end slope of $1.64$ and  a bright
end slope of $3.43$ ($z<3$) or $2.58$ ($z>3$). 

\begin{figure}
\epsscale{0.6}
\plotone{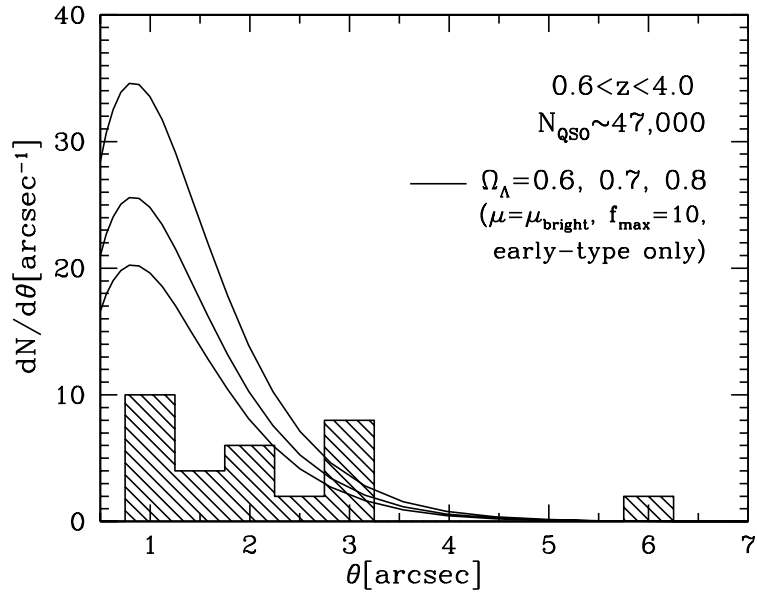}
\caption{Tentative comparison of image separation distributions of
 lensed quasars in the current SDSS quasar sample. Lines show the
 theoretical expectations (see text for details of computation) using an
 SIS model, for different values of the cosmological constant,
 $\Omega_\Lambda=0.6$, $0.7$, and $0.8$ from bottom to top,
 respectively. Shaded region indicates the distribution of all lensed
 quasars discovered and recovered in the SDSS (see Table
 \ref{table:lens}). The large-separation lensed quasar SDSS J1004+4112
 ($\theta=14\farcs6$) is not included in this plot. \label{fig:sepdist}}
\end{figure}

The result is shown in Figure \ref{fig:sepdist}. As seen, the number of
$\sim 3''$ lenses already exceeds the theoretical expectations, when we
adopt the standard value of the cosmological constant, 
$\Omega_\Lambda\sim 0.7$. The situation is similar if we
increase $\Omega_\Lambda$ to $0.8$. In terms of the constraint on the
cosmological constant, the excess in the bin centered at $\theta=3''$ 
is described as $\Omega_\Lambda>0.90$ (68\% confidence limit) assuming a
Poisson distribution, which is highly inconsistent with recent
measurements of the cosmological constant $\Omega_\Lambda\sim 0.70$ with
errors of less than 10\% \citep[e.g.,][]{tegmark04}. We emphasis that
our current lens sample is quite incomplete; indeed, currently we have
several $\sim 3''$ quasar pairs with the same redshifts which could also
be gravitational lenses \citep{hennawi04}, which need deep imaging to
find the putative lens galaxies. Thus observations of these quasar pairs
as well as lens candidates should be conducted, which might make the
discrepancy even larger.  Such an excess may indicate that our simple
treatment of lensing statistics is inaccurate and that we need to
include other effects.   

\section{Summary}
\label{sec:sum}

We report the discovery of two gravitationally lensed quasars, SDSS
J1001+5027 and SDSS J1206+4332. The systems were identified as new lens
candidates in the SDSS, and confirmed as lenses by spectroscopic and
imaging observations at the ARC 3.5m and the UH88 telescope. SDSS
J1001+5027 is a lensed quasar at $z=1.838$, and consists of two lensed
images separated by $2\farcs86$. SDSS J1206+4332 is a lensed quasar at
$z=1.789$, and consists of two lensed images separated by
$2\farcs90$. In each system we have identified the galaxy responsible
for the lensing.

We have found that the lens systems are complicated. The imaging data
clearly show other galaxies that are very close to the main lens
galaxies. The lens modeling has shown that these galaxies affect the lens
potentials significantly. We have examined the wide field images, which
show an enhancement of the galaxy number density within $60''$ of SDSS
J1001+5027. Spectroscopic follow-up observations are needed to see if
the group/cluster ($z\sim 0.2$) is actually associated with the lens 
galaxy.  

Although the SDSS lens survey is ongoing, we have made a preliminary
comparison of theoretical lensing probability distributions with
the observed distribution. We have found that the number of lenses with 
$\theta\sim 3''$ already exceeds the theoretical expectations. We still
have many lens candidates with $\theta\gtrsim 3''$ that remain to be
observed \citep{hennawi04}; if some of them turn out to be true
gravitational lenses, the conflict will become even stronger. This excess
may be caused by external convergence and shear fields which we have not
taken into account in our calculation. Basically, external shear 
broadens the distribution of image separations for a given mass of a lens   
object \citep[e.g.,][]{huterer05}. This broadening is enough to enhance 
the lensing probability at $\theta\gtrsim 3''$ because the lensing 
probability at the image separation region is a strong function of image separations.
More significant enhancement may be achieved by external convergence,
since it increases both image separation and lensing probability.
Indeed, among five intermediate-separation lenses in Table
\ref{table:lens}, two lens systems (RX J0911+0551 and Q 0957+561) lie in
clusters, and two other lens systems (SDSS J0903+5028 and SDSS
J1001+5027) also appear to lie in dense environments. In addition, the
two lens systems reported in this paper are complex in the sense that
galaxies very close to the main lens galaxies affect the lens
potentials. Other possible systematic effects include triaxiality of
lenses \citep{oguri04c} and massive substructures \citep{cohn04}.
In either case, lens statistics at this image separation should
be done with caution; simple models that consider only isolated single
lens objects can be misleading. 

\acknowledgments
We thank Paul Schechter for useful comments, and anonymous referee for
many suggestions. M.~O. and N.~I. are supported by JSPS through JSPS
Research Fellowship for Young Scientists. J.~F.~H is currently supported
by NASA through Hubble Fellowship grant  \#01172.01-A awarded by the Space
Telescope Science Institute, which is operated by the Association of
Universities for Research in Astronomy, Inc., for NASA, under contract
NAS 5-26555. 

Funding for the creation and distribution of the SDSS Archive has been
provided by the Alfred P. Sloan Foundation, the Participating
Institutions, the National Aeronautics and Space Administration, the
National Science Foundation, the U.S. Department of Energy, the
Japanese Monbukagakusho, and the Max Planck Society. The SDSS Web site
is http://www.sdss.org/. 

The SDSS is managed by the Astrophysical Research Consortium (ARC) for
the Participating Institutions. The Participating Institutions are The
University of Chicago, Fermilab, the Institute for Advanced Study, the
Japan Participation Group, The Johns Hopkins University, the Korean
Scientist Group, Los Alamos National Laboratory, the
Max-Planck-Institute for Astronomy (MPIA), the Max-Planck-Institute
for Astrophysics (MPA), New Mexico State University, University of
Pittsburgh, Princeton University, the United States Naval Observatory,
and the University of Washington. 

This work is based on observations obtained with the Apache Point
Observatory 3.5-meter telescope, which is owned and operated by the
Astrophysical Research Consortium, and with the University of Hawaii 
2.2-meter telescope.

\clearpage

\begin{deluxetable}{cccccccc}
\rotate
\tablewidth{0pt}
\tablecaption{SDSS Astrometry and Photometry\label{table:sdss}}
\tablehead{\colhead{Object} & \colhead{R.A. (J2000.0)} & \colhead{Dec. (J2000.0)} & \colhead{$u$} & \colhead{$g$} &
 \colhead{$r$} & \colhead{$i$} & \colhead{$z$}} 
\startdata
\multicolumn{8}{c}{SDSS~J1001+5027} \vspace*{1.5mm} \\ \hline \vspace*{-2.0mm} \\ 
A & 10 01 28.61 & +50 27 56.9 & $17.71\pm0.02$ & $17.60\pm0.02$ & $17.55\pm0.03$ & $17.36\pm0.03$ & $17.33\pm0.04$ \\
B & 10 01 28.35 & +50 27 58.5 & $18.60\pm0.08$ & $18.35\pm0.08$ & $18.11\pm0.06$ & $17.71\pm0.06$ & $17.62\pm0.05$ \\
\cutinhead{SDSS~J1206+4332}
A & 12 06 29.65 & 43 32 17.6 & $18.66\pm0.03$ & $18.80\pm0.05$ & $18.75\pm0.05$ & $18.53\pm0.04$ & $18.48\pm0.04$ \\
B & 12 06 29.65 & 43 32 20.6 & $19.69\pm0.08$ & $19.43\pm0.12$ & $19.33\pm0.08$ & $19.22\pm0.08$ & $18.83\pm0.04$ \\
\enddata
\tablecomments{PSF magnitudes returned by the SDSS photometric pipeline
 are shown. Reddening corrections are not applied.} 
\end{deluxetable}
\begin{deluxetable}{cccccc}
\tablewidth{0pt}
\tablecaption{UH88 Astrometry and Photometry\label{table:uh88}} 
\tablehead{\colhead{Object} & \colhead{$x$[arcsec]\tablenotemark{a}} &
 \colhead{$y$[arcsec]\tablenotemark{a}} & 
 \colhead{$V$\tablenotemark{b}} & \colhead{$R$\tablenotemark{b}} &
 \colhead{$I$\tablenotemark{b}}}  
\startdata
\multicolumn{6}{c}{SDSS~J1001+5027} \vspace*{1.5mm} \\ \hline \vspace*{-2.0mm} \\ 
A  & $0.000\pm0.005$ & $0.000\pm0.005$ & $18.12\pm0.01$ & $17.69\pm0.01$ & $17.32\pm0.01$ \\
B  & $2.418\pm0.009$ & $1.526\pm0.005$ & $18.53\pm0.01$ & $17.98\pm0.01$ & $17.48\pm0.01$ \\
G1 & $1.779\pm0.049$ & $0.857\pm0.123$ & \nodata & $20.51\pm0.03$ & $19.63\pm0.03$ \\
G2 & $1.795\pm0.088$ & $-0.700\pm0.053$ & \nodata & $20.91\pm0.04$ & $20.15\pm0.03$ \\
G1+G2\tablenotemark{c} & \nodata & \nodata & $22.09\pm0.08$ & \nodata & \nodata \\
\cutinhead{SDSS~J1206+4332}
A  & $0.000\pm0.011$  & $0.000\pm0.010$ & $18.63\pm0.02$ & $18.54\pm0.01$ & $18.05\pm0.02$ \\
B  & $-0.098\pm0.006$ & $2.894 \pm0.009$ & $19.12\pm0.02$ & $18.95\pm0.02$ & $18.38\pm0.02$ \\
G1 & $-0.664\pm0.137$ & $1.748\pm0.028$ & \nodata & $21.34\pm0.07$ & $19.51\pm0.03$ \\
G2 & $1.320\pm0.147$  & $5.999\pm0.148$ & $22.43\pm0.13$ & $21.16\pm0.08$ & $19.63\pm0.04$ \\
G3 & $-2.052\pm0.200$  & $2.397\pm0.152$ & $21.78\pm0.09$ & $22.05\pm0.10$ & $22.57\pm0.11$ \\
\enddata
\tablenotetext{a}{Positions relative to the brighter quasar components
 are presented. The positive directions of $x$ and $y$ are defined by
 West and North, respectively. Error bars do not include the error of
 the image scale. }
\tablenotetext{b}{Error bars do not include the zero-point error of $\sim 0.1$ mag.}
\tablenotetext{c}{In the $V$-band image it was difficult to decompose G1
 and G2 because of the faintness if the image, thus we show the total magnitude
 of these two galaxies.}
\end{deluxetable}
\begin{deluxetable}{ccccc}
\tablewidth{0pt}
\tablecaption{Lens modeling\label{table:model}}
\tablehead{\colhead{Model} & \colhead{$R_{\rm E}$[arcsec]} &
 \colhead{$e$ or $\gamma$} &
 \colhead{$\theta_e$ or $\theta_\gamma$[deg]\tablenotemark{a}} & \colhead{$\Delta
 t$[$h^{-1}$day]\tablenotemark{b}}} 
\startdata
\multicolumn{5}{c}{SDSS~J1001+5027} \vspace*{1.5mm} \\ \hline \vspace*{-2.0mm} \\ 
SIS$+$shear & $1.35$ & $0.09$ & $11.9$ & $42.4$ \\
SIE         & $1.38$ & $0.25$ & $10.9$ & $47.0$ \\
\cutinhead{SDSS~J1206+4332}
SIE+G2\tablenotemark{c}   & $1.39$ & $0.31$ & $-89.3$ & $92.6$ \\
\enddata
\tablecomments{See \S \ref{sec:model} for details of the lens models
 adopted here.}
\tablenotetext{a}{Each position angle is measured East of North.}
\tablenotetext{b}{For the redshifts of lens galaxies, we assumed a
 lens redshift of $z=0.3$ for SDSS J1001+5027, and the redshift
 of strong \ion{Mg}{2} absorption, $z=0.748$, for SDSS~J1206+4332.
We also assumed $\Omega_M=0.3$ and $\Omega_\Lambda=0.7$, but the values
 are quite insensitive to these cosmological parameters.}
\tablenotetext{c}{The Einstein radius of G2, $R_{\rm E}$(G2), is fixed
 to $1''$.} 
\end{deluxetable}
\begin{deluxetable}{crcccr}
\tablewidth{0pt}
\tablecaption{Gravitationally Lensed Quasars (Re-)Discovered in the
 SDSS: Current Status\label{table:lens}}  
\tablehead{\colhead{Name} & \colhead{$\theta_{\rm max}$\tablenotemark{a}} &
 \colhead{$z_s$} & \colhead{$z_l$} & 
 \colhead{Comments} & \colhead{Ref.}}
\startdata
Q 0142$-$100     & $2\farcs22$ & $2.72$ & $0.49$ & previously known lens & 1\\
SDSS J0246$-$0825& $1\farcs04$ & $1.69$ & $0.72?$& & 2\\
SDSS J0903+5028  & $2\farcs83$ & $3.58$ & $0.39$ & targeted as a galaxy& 3\\
RX J0911+0551    & $3\farcs25$ & $2.80$ & $0.77$ & previously known lens & 4\\
SBS 0909+523     & $1\farcs11$ & $1.38$ & $0.83$ & previously known lens & 5\\
SDSS J0924+0219  & $1\farcs78$ & $1.52$ & \nodata& & 6\\
Q 0957+561       & $6\farcs17$ & $1.41$ & $0.36$ & previously known lens & 7\\
SDSS J1001+5027  & $2\farcs86$ & $1.84$ & \nodata& & 8\\
SDSS J1004+4112  & $14\farcs62$& $1.73$ & $0.68$ & large-separation lens & 9\\
SDSS J1021+4913  & $1\farcs05$ & $1.72$ & \nodata& & 10\\
PG 1115+080      & $2\farcs43$ & $1.72$ & $0.31$ & previously known lens & 11\\
SDSS J1138+0314  & $1\farcs46$ & $2.44$ & \nodata& & 12\\
SDSS J1155+6346  & $1\farcs83$ & $2.89$ & $0.18?$& & 13\\
SDSS J1206+4332  & $2\farcs90$ & $1.79$ & $0.75?$& & 8\\
SDSS J1226$-$0006& $1\farcs24$ & $1.13$ & $0.52?$& & 14\\
SDSS J1335+0118  & $1\farcs56$ & $1.57$ & \nodata& & 15\\
SDSS J1650+4251  & $1\farcs18$ & $1.55$ & \nodata& & 16\\
\enddata
\tablecomments{The SDSS strongly lensed quasar survey is still ongoing,
and we have many other candidates requiring follow-up observations. 
Therefore, this list is far from complete. This list includes only
lenses that have SDSS spectra.}
\tablenotetext{a}{The maximum separation between multiple images.} 
\tablerefs{(1) \citealt{surdej87}; (2) \citealt{inada05b}; 
(3) \citealt{johnston03}; (4) \citealt{bade97}; (5) \citealt{oscoz97}; 
(6) \citealt{inada03a}; (7) \citealt{walsh79}; (8) this paper; 
(9) \citealt{inada03b}; (10) \citealt{pindor05}; (11) \citealt{weymann80}; 
(12) \citealt{burles05}; (13) \citealt{pindor04}; (14) \citealt{inada05a}; 
(15) \citealt{oguri04b}; (16) \citealt*{morgan03}}
\end{deluxetable}
\clearpage
\clearpage
\clearpage

\end{document}